\documentclass[a4paper,11pt]{article}
\usepackage{pos}

\title{AstroSat View of Blazar OJ 287: A complete evolutionary cycle of
HBL Component from end-phase to disappearance and Re-emergence}
 \ShortTitle{AstroSat View of BL Lac OJ 287}

\author*[a,\dag]{Pankaj Kushwaha}
\notes{\note{Aryabhatta Postdoctoral Fellow}}

\author[b]{K. P. Singh}
\author[c,d]{A. Sinha}
\author[e]{Main Pal}
\author[d]{G. Dewangan}
\author[f]{A. Agarwal}

\affiliation[a]{Aryabhatta Research Institute of Observational Sciences
(ARIES), Nainital 263001, India}

\affiliation[b]{Indian Institute of Science Education and Research Mohali, Knowledge City, Sector 81, SAS Nagar, Punjab 140306, India}

\affiliation[c]{Laboratoire Univers et Particules de Montpellier, Université de Montpellier, CNRS/IN2P3, CC 72, Place Eugène Bataillon, F-34095 Montpellier Cedex 5, France}

\affiliation[d]{Inter-University Centre for Astronomy and Astrophysics, Ganeshkhind, Pune 411 007, India}

\affiliation[e]{Centre for Theoretical Physics, Jamia Millia Islamia, New Delhi-110025, India}

\affiliation[f]{Raman Research Institute, C. V. Raman Avenue, Sadashivanagar, Bengaluru 560080, India}

\emailAdd{pankaj.kushwaha@aries.res.in}

\abstract{We report three AstroSat observations of BL Lacertae object OJ 287.
The three observations caught it in very different flux states that are connected 
to different broadband spectral states. These observations trace the source spectral
evolution from the end-phase of activity driven by a new, additional HBL like
emission component in 2017 to its complete disappearance in 2018 and re-emergence
in 2020. The 2017 observation shows a comparatively flatter optical-UV and X-ray 
spectrum. Supplementing it with the simultaneous NuSTAR monitoring indicates a 
hardening at the high-energy-end. The 2018 observation shows a harder X-ray spectrum 
and a sharp decline or cutoff in the optical-UV spectrum, revealed thanks to the Far-UV 
data from AstroSat. The brightest of all, the 2020 observation shows a hardened 
optical-UV spectrum and an extremely soft X-ray spectrum, constraining the low-energy 
peak of spectral energy distribution at UV energies – a characteristic of HBL blazars. 
The contemporaneous MeV-GeV spectra from LAT show the well-known OJ 287 spectrum during 
2018 but a flatter spectrum during 2017 and a hardening above ~1 GeV during 2020. 
Modeling broadband SEDs show that 2018 emission can be reproduced with a one-zone 
leptonic model while 2017 and 2020 observations need a two-zone model, with the 
additional zone emitting an HBL radiation.}

\FullConference{37$^{\rm{th}}$ International Cosmic Ray Conference (ICRC 2021)\\
		July 12th -- 23rd, 2021\\
		Online -- Berlin, Germany}

\begin{document}
\maketitle

\section{Introduction}
OJ 287 is a BL Lacartae (BL Lac) object located at a cosmological redshift
of 0.306 \citep{1985PASP...97.1158S,2010A&A...516A..60N}. The BL Lac designation 
was originally coined for sources with very weak or completely absent emission
line features but showed strong and rapid variations in continuum flux, polarization,
and have an inverted radio spectrum. The source has been extensively studied and shows
an entirely jet-dominated continuum spanning the entire accessible electromagnetic 
spectrum from radio to GeV/TeV gamma-rays \citep[e.g.][and references therein]
{2010ApJ...716...30A,2013MNRAS.433.2380K,2018ApJ...863..175G}. 

The broadband emission from OJ 287 shows the characteristic double-humped spectral
energy distribution of blazars with a low-energy peak around near-infrared (NIR)
bands and a high energy peak around $\rm\sim$ 100 MeV \citep[e.g.][]{2010ApJ...716...30A,2013MNRAS.433.2380K}. In terms of SED based classification scheme, OJ 287 is a
low-energy-peaked (LBL) blazar. Modeling of SEDs constructed using simultaneous/contemporaneous data with constraints from observations in different energy
bands show that the MeV-GeV gamma-ray emission is due to external Comptonization (EC) of
a $\rm\sim 250$ K thermal photon field while X-ray is due to synchrotron self-Compton 
\citep[SSC;][]{2013MNRAS.433.2380K}. Though blazars show flux variability overall  observationally feasible timescales, significant spectral evolution indicating
a new emission component or change of spectral sequence are rare \citep[e.g.][references therein]{2020arXiv201014431K}. For OJ 287, however, multi-wavelength observations show that OJ 287 is a spectrally very dynamic blazar with strong
spectral changes in all energy bands for a much-extended duration \citep[e.g.][and references therein]{2017ICRC...35..650B,2017IAUS..324..168K,2018MNRAS.473.1145K,
2018MNRAS.479.1672K,2020ApJ...890...47P,2020MNRAS.498L..35K}.

OJ 287 is been exhibiting continuous multi-wavelength activity in phases,
one followed by the next, since the end-2015 \citep[e.g.][]{2017MNRAS.465.4423G,2020ATel13785....1K,2020MNRAS.498L..35K}, with each activity phase being
spectrally distinct compared to the preceding one \citep[e.g.][]{2017ICRC...35..650B,2017IAUS..324..168K,2018MNRAS.473.1145K,2018MNRAS.479.1672K,2018MNRAS.480..407K,2020MNRAS.498L..35K,2020ApJ...890...47P,2020arXiv201014431K,2021arXiv210504028P,Singh21}. MW spectro-temporal and spectral studies of the 
first phase ($\rm\sim$ end-2015 -- mid-2016) of activity discovered a distinct NIR-optical spectrum with a break
consistent with a thermal feature, a hardening of the\ MeV-GeV spectrum 
with a shift in the location of high-energy hump \citep{2018MNRAS.473.1145K},
as well as the discovery of a soft X-ray excess for the first time in any BL Lac
object \citep{2020ApJ...890...47P}. This was followed by an intense and bright
X-ray to optical flux activity phase ($\rm\sim$ mid-2016 -- mid-2017), driven by an extremely soft X-ray spectrum
\citep{2017IAUS..324..168K} with the resultant broadband SED similar to a combined
presence of the typical LBL SED of OJ 287  and an additional high-energy-peaked (HBL)
like emission component \citep{2018MNRAS.479.1672K} but with some peculiar timing
features \citep{2018MNRAS.479.1672K,2020MNRAS.498L..35K}. During this phase, VERITAS
reported its first ever activity at very high energies \citep[VHEs;][]{2017ICRC...35..650B}. With the weakening of this new component, 
the source slowly returned to its typical LBL spectral state but only for a 
comparatively short duration of $\rm\sim 50$ days but the HBL-like component
slowly and steadily gained prominence \citep{2020arXiv201014431K}, leading to
another activity with a very
soft X-ray spectrum in 2020 \citep{2020MNRAS.498L..35K}.

In this proceeding, we report our MW spectral analysis and modeling of the long-exposure observations of 
OJ 287 by {\it AstroSat} taken respectively in 2017, 2018, and 2020 that correspond
to the different X-ray flux states of the source during different activity phases
mentioned above. The 2017 is around the end-phase of the HBL drive activity
phase of 2016--2017, 2018 is around the period when this additional component has
started to gain strength after disappearance and the 2020 is after the peak of re-emerged HBL driven activity phase. In the next section (\S\ref{sec:Obs}), we provide a brief
overview of {\it AstroSat} observations and the results of our spectral analysis. In \S \ref{sec:discuss}, we discuss our findings in the 
context of jet physics and conclude with a summary in \S \ref{sec:Sum}.

\section{AstroSat Observations}\label{sec:Obs}
{\it AstroSat} is the first Indian multi-wavelength facility launched in 2015
\citep{2014SPIE.9144E..1SS} with four data acquisition payloads for simultaneous astronomical observations
-- the Ultraviolet Imaging Telescope \citep[UVIT;][]{2017AJ....154..128T}, Soft
X-ray Telescope \citep[SXT;][]{2016SPIE.9905E..1ES}, Large Area X-ray Proportional
Counter \citep[LAXPC;][]{2017ApJS..231...10A}, and  Cadmium Zinc Telluride Imager \citep[CZTI;][]{2017JApA...38...31B}. Together these provide simultaneous coverage
of the electromagnetic spectrum from optical to hard X-rays. The UVIT and SXT
have focusing optics while LAXPC and CZTI are open detectors.

{\it AstroSat} targeted OJ 287 on three occasions -- in 2017, 2018, and 2020 respectively with the source being in 3 different X-ray flux states. We explored
these observations and found useful data only in UVIT and SXT. Data from the 
other two -- LAXPC and CZTI were noise-dominated and thus, are not considered.

\begin{figure}
\includegraphics[width=1.\textwidth]{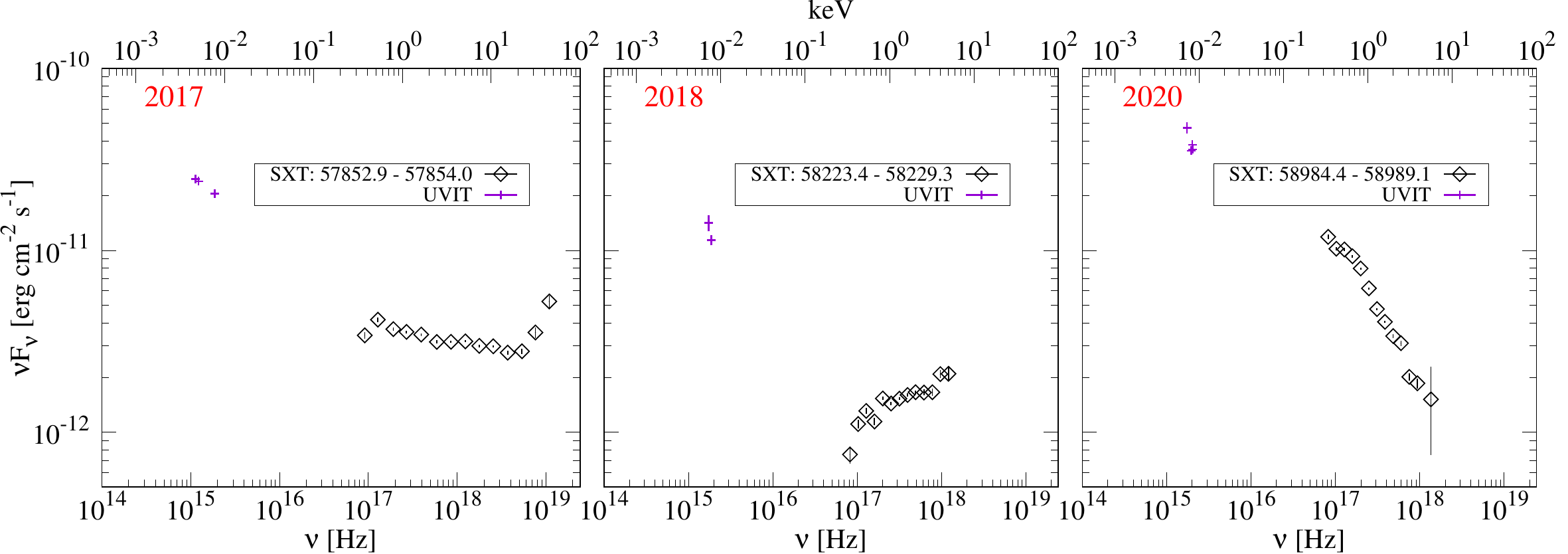}
\caption{Multi-wavelength SEDs of OJ 287 from the three long-exposure observations
with {\it AstroSat} in 2017, 2018, and 2020 respectively.}
\label{fig:aSatSED}
\end{figure}

For data reduction, we followed the standard recommended prescription 
for UVIT and SXT, and detailed treatment is presented in \citep{Singh21}. 
We found no appreciable variability in the SXT
and UVIT\footnote{FUV/NUV data show variations at a level of few percent that
has negligible effect on overall SED.} light curves. The X-ray spectrum of all three is consistent with a power-law spectrum (\(N(E)
\sim E^{-\Gamma}\)) with a photon spectral index, $\rm\Gamma$ of $\rm 2.08\pm0.03$,
$\rm 1.82\pm0.06$, and $\rm 2.5\pm0.04$ respectively
for 2017, 2018, and 2020 observations assuming a fixed Galactic neutral hydrogen
column density of $\rm 2.4\times10^{20}~cm^{-2}$. 
Figure \ref{fig:aSatSED} shows the MW SEDs constructed using the {\it AstroSat}
data. The X-ray spectrum of 2017 also includes the X-ray data from the {\it NuSTAR}
facility.

%

\section{Discussion}\label{sec:discuss}
\begin{figure}
\includegraphics[width=1.\textwidth]{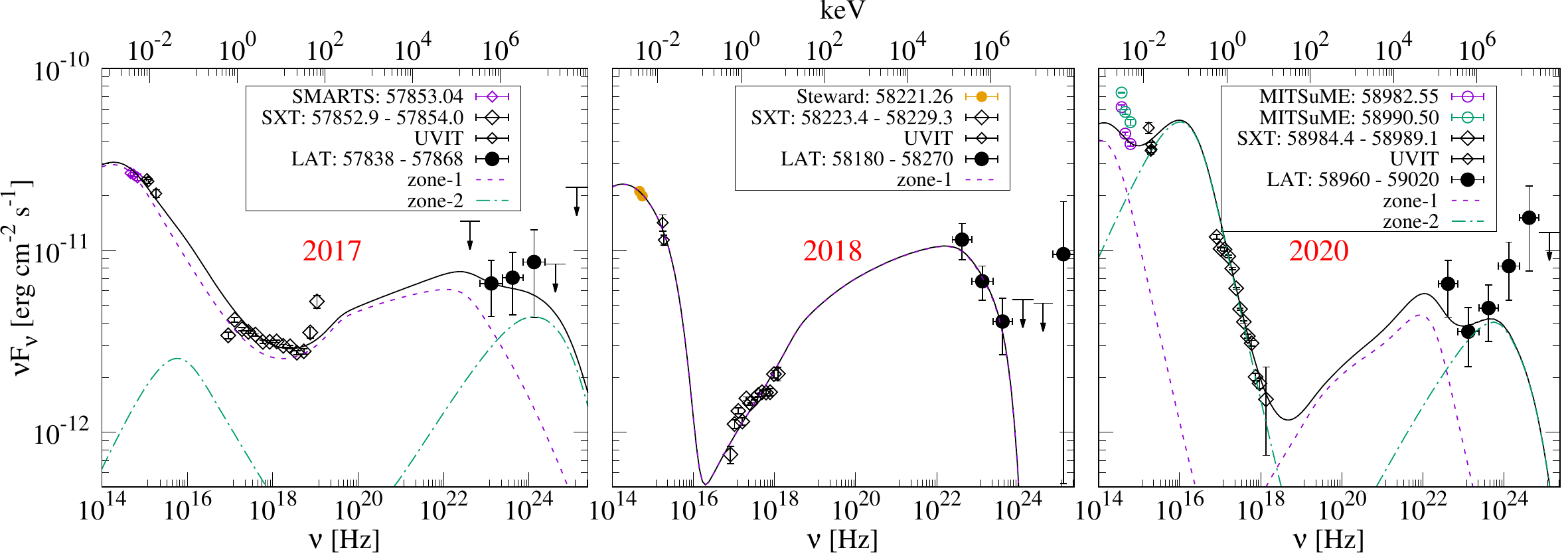}
\caption{Broadband SEDs of OJ 287 corresponding to the three {\it AstroSat} 
observations. }
\label{fig:mwSEDs}
\end{figure}

We found that the three {\it AstroSat} observations performed during three different
X-ray (0.3 -- 10 keV) flux states of the source actually correspond to three distinct
X-ray spectral phases of the source. Together the X-ray spectra shown in figure \ref{fig:aSatSED} covers the entire
range of spectral shape seen in blazars, from a hard -- the 2018 spectrum,
intermediate -- the 2017 spectrum, to an extremely soft -- the 2020 spectrum. The
corresponding broadband SEDs shown in figure \ref{fig:mwSEDs} shows spectral changes
at UV and MeV-GeV gamma-rays (from {\it Fermi}-LAT; \citep{2009ApJ...697.1071A}), indicating the spectral changes to a broadband emission
component. Note that the source is too weak 
at MeV-GeV energies over the {\it AstroSat} monitoring periods and thus, a 
much longer duration data is used to extract the gamma-ray spectra. Any strong
spectral changes are expected to be preserved though the intensity may be reduced
substantially. Our focus here is overall broadband spectral changes and not 
its level and thus, extraction using data from a longer duration is perfectly fine.

To explore the possible reason for the primary driver of the X-ray 
spectrum state, we modeled the 2018 and 2020 SEDs as the two-ends
of the overall spectral evolution sequence. The NUV-FUV data during 2018 
show a sharp softening/cutoff at the high-energy-end of the optical-UV spectrum.
We found that a one-zone model with emission resulting from a relativistic
electron population emitting synchrotron, SSC, and EC of a $\sim 250$ K blackbody
photon field as proposed in \citep{2013MNRAS.433.2380K} can satisfactorily reproduce the 2018 broadband SED. For the 2020
SED, the hardening seen at UV energies with respect to the optical
and a similar trend at MeV-GeV energies is inconsistent with the one-zone
scenario employed for 2018 SED and indicates a new emission component.
A power-law extrapolation of the X-ray spectrum to UV energies indicates a peak around
UV energies -- a characteristic of the HBL emission component. Thus, we modeled
the 2020 SED with a two-zone model \citep[e.g.][]{2018MNRAS.479.1672K,Singh21}
with the additional zone emitting an HBL spectrum incorporating only synchrotron (optical to X-rays) and SSC (gamma-rays) as typically invoked for HBL blazars
and found it sufficient to reproduce the 2020 broadband SED.

The above modeling for 2018 and 2020 SEDs presents two plausible scenarios for the
X-ray spectrum of 2017 (intermediate between 2018 and 2020 X-ray spectra) -- a continuation of optical-UV synchrotron spectrum
to the X-ray band, supported by the NUV-FUV data or a much weaker HBL component. We
found that the intermediate X-ray spectrum is primarily due to the continuation of the 
synchrotron to X-ray energies. A significantly weaker (by a factor of $\sim 4$)
contribution of the HBL component with respect to the LBL component is needed to
reproduce the observed X-ray spectrum but at MeV-GeV, its contribution is significant.
These modeled spectra for all the three SEDs are shown in figure \ref{fig:mwSEDs}
(ref \citep{Singh21} for details).

\section{Summary}\label{sec:Sum}
We presented a spectral analysis of the three long-exposure observations of OJ 287
with  {\it AstroSat} taken during three distinct X-ray flux states and found these
to be spectrally distinct too. Together these X-ray spectra covers the whole range of
spectral phases exhibited by the blazar class -- from a hard in 2018, intermediate
state in 2017, to an extremely soft X-ray spectrum in 2020. 

The NUV-FUV data of 2018 show a spectral sharpening/cutoff in the high-energy-end
of the optical-UV spectrum and a one-zone leptonic emission model can reproduce
the broadband SED fairy well. For the 2020 SED, an additional emission zone emitting
an HBL spectrum is needed while for 2017, the X . The HBL component has almost
negligible contribution at X-ray energies but contributes significantly at MeV-GeV
energies, indicating that in the LBL state, the X-ray spectral changes are primarily
driven by the evolution of the high-energy-end of the optical-UV synchrotron spectrum.

\acknowledgments
PK acknowledges financial support from ARIES Aryabhatta Postdoctoral Fellowship (A-PDF)
grant (AO/A-PDF/770). 

\bibliographystyle{JHEP}
\bibliography{icrc2021}


%
%
%

\end{document}